%% file: 0-main.tex
\documentclass[conference,10pt]{IEEEtran}
\IEEEoverridecommandlockouts
\input{00-packages}
\input{000-macros}
\title{Distributed Quantum Computation with Minimum Circuit Execution Time over Quantum Networks \vspace*{-0.1in}}

\author{\IEEEauthorblockN{Ranjani G. Sundaram}
\IEEEauthorblockA{
\textit{ Stony Brook University, NY}}
\and
\IEEEauthorblockN{Himanshu {Gupta}}
\IEEEauthorblockA{
\textit{ Stony Brook University, NY}}
\and
\IEEEauthorblockN{C. R. Ramakrishnan}
\IEEEauthorblockA{
\textit{ Stony Brook University, NY}}}
\IEEEaftertitletext{\vspace{-3\baselineskip}}
\begin{document}
\maketitle
\thispagestyle{plain}
\pagestyle{plain}

\begin{abstract}
Present quantum computers are constrained by limited qubit capacity and restricted physical connectivity, leading to challenges in large-scale quantum computations.
Distributing quantum computations across a network of quantum computers is a promising
way to circumvent these challenges and facilitate large quantum computations. 
However, distributed quantum computations require entanglements (to execute remote
gates) which can incur significant generation latency and, thus, lead to decoherence
of qubits. 
In this work, we consider the problem of distributing quantum circuits across a quantum network to minimize the execution time. The problem entails mapping the circuit qubits to network memories, 
including within each computer since limited connectivity within computers can affect the circuit
execution time. 

We provide two-step solutions for the above problem: In the first step, we allocate qubits to memories to minimize the estimated execution time; for this step, we design an efficient
algorithm based on an approximation algorithm for the max-quadratic-assignment problem. In the second step, we determine an efficient execution scheme, including generating required entanglements with minimum latency under the network resource and decoherence constraints; for this step, we develop two algorithms with appropriate performance guarantees under certain settings or assumptions.
We consider multiple protocols for executing remote gates, viz., telegates and cat-entanglements. 
With extensive simulations over NetSquid, a quantum network simulator, we demonstrate the effectiveness of our developed techniques and show that they outperform a scheme based on prior work by up to 95\%.
\end{abstract}
\input{1-Introduction}

\input{2-Background}
\input{3-Problem}

\input{5-Step-1}
\input{6-Step-2}
\input{7a-Cat}

\input{7b-other-gens}
\input{8-Evaluation}
\input{9-Conclusion}
\newpage
\bibliographystyle{plainurl}
\begin{small}
\bibliography{ref}
\end{small}
\end{document}

%% file: 00-packages.tex
\usepackage{algorithm}
\usepackage{algorithmic}
\usepackage{amsfonts}
\usepackage{amsmath}
\usepackage{mathrsfs}
\usepackage{color}
\usepackage{esvect}
\usepackage{gensymb}
\usepackage{graphicx}
\usepackage{mathtools}
\usepackage[binary-units=true]{siunitx}
\usepackage{stfloats}
\usepackage{url}
\usepackage{xspace}
\usepackage{wrapfig}
\usepackage[normalem]{ulem}
\usepackage{subcaption}
\usepackage[shortlabels]{enumitem}

\makeatletter
\newcommand\fs@spaceruled{\def\@fs@cfont{\bfseries}\let\@fs@capt\floatc@ruled
  \def\@fs@pre{\vspace{0.5\baselineskip}\hrule height.8pt depth0pt \kern2pt}%
  \def\@fs@post{\kern2pt\hrule\relax\vspace{-0.4cm}}%
  \def\@fs@mid{\kern2pt\hrule\kern2pt}%
  \let\@fs@iftopcapt\iftrue}
\makeatother
\floatstyle{spaceruled}
\restylefloat{algorithm}
\usepackage[font=small]{caption}
\usepackage[sort,square,numbers]{natbib}

%% file: 000-macros.tex
\newcommand{\eat}[1]{}

\newcommand{\blue}[1]{{\color{black}#1}}

\newcommand{\cblue}{\color{black}}
\newcommand{\cblack}{\color{black}}

\newcommand{\para}[1]{\smallskip \noindent {\bf #1}}
\newcommand{\softpara}[1]{\smallskip \noindent \underline{#1}}

\newcommand{\showOutline}{s}
\newcommand{\hide}[1]{}

\definecolor{darkgreen}{rgb}{0,0.5,0}
\definecolor{darkblue}{rgb}{0,0,0.5}

\newcommand{\outline}[2]{\if#1\showOutline {\color{darkgreen}#2} \fi}
\newcommand{\detailedoutline}[2]{\if#1\showOutline {\color{darkblue}#2} \fi}

\newcommand{\todo}[2]{\if#1\showOutline {#2} \fi}

\newcommand{\purp}[1]{} 

\def\blackbox{\hfill {\vrule height6pt width6pt depth0pt}}
\newcommand{\vsplem}{\vspace{0.025in}}

\newtheorem{lem}{Lemma}
\newtheorem{thm}{Theorem}
\newtheorem{obv}{Observation}

\newenvironment{lem-wo-prf}{\begin{lem} \nopagebreak}{{\hfill$\blackbox$} \end{lem}}
\newenvironment{lem-w-prf}{\begin{lem} \nopagebreak}{\end{lem}}
\newenvironment{thm-w-prf}{\begin{thm} \nopagebreak}{\end{thm}}
\newenvironment{thm-wo-prf}{\begin{thm} \nopagebreak}{{\hfill$\blackbox$} \end{thm}}
\newenvironment{prf}{{\sc Proof:} \nopagebreak }{{\hfill$\blackbox$}}

\newcounter{packednmbr}

\newcommand{\ges}{{\tt GenerateEPs}\xspace}
\newcommand{\ged}{{\tt GED}\xspace}
\newcommand{\EP}{EP\xspace}

\newcommand{\es}{ES\xspace}

\newcommand{\EPs}{EPs\xspace}

\newcommand{\E}{\mbox{${\cal E}$}\xspace}
\newcommand{\C}{\mbox{${\cal C}$}\xspace}

\newcommand{\id}[1]{\textit{#1}}
\newcommand{\cz}{\id{CZ}\xspace} 

\newcommand{\dqc}{{\tt DQC}\xspace}
\newcommand{\dqcqr}{{\tt DQC-QR}\xspace}
\newcommand{\qr}{{\tt QR}\xspace}

\newcommand{\minqap}{min-{\tt QAP}\xspace}
\newcommand{\maxqap}{max-{\tt QAP}\xspace}

\newcommand{\greedy}{{\tt Greedy'}\xspace}
\newcommand{\greedytg}{{\tt Greedy-TG}\xspace}
\newcommand{\greedyg}{{\tt Greedy}\xspace}
\newcommand{\greedycat}{{\tt Greedy-CE}\xspace}
\newcommand{\DPcat}{{\tt DP-CE}\xspace}
\newcommand{\DP}{{\tt DP-TG}\xspace}
\newcommand{\CE}{CE\xspace}

\newcommand{\ces}{{\tt CESelection}\xspace}
\newcommand{\dsvc}{{\tt DSVC}\xspace}

\newcommand{\Caleffi}{{\tt Disjoint-Paths}\xspace}

\newcommand{\CNOT}{\ensuremath{\mathit{CNOT}}\xspace}
\newcommand{\CZ}{\ensuremath{\mathit{CZ}}\xspace}

\newcommand{\qi}{\ensuremath{q_i}\xspace} 
\newcommand{\qj}{\ensuremath{q_j}\xspace}

\newtheorem{defin}{Definition}

\newenvironment{theorem}{\begin{thm} \nopagebreak}{{\hfill$\blackbox$} \end{thm}}

%% file: 1-Introduction.tex
\section{\bf Introduction}
\label{sec:intro}


Present quantum computers have limited qubit capacity and are susceptible to noise due to decoherence and gate operations. Error-correcting codes can be used to overcome noisy gate operations, but that results in a blowup in the number of qubits, which further exacerbates 
the qubit capacity hurdle. 
Distributing quantum computations over a network of quantum computers (QCs) is a promising way 
to facilitate large computations over current QCs. 
However, distributed quantum computations require entanglements (to execute remote
gates) which can incur significant generation latency and, thus, lead to decoherence
of qubits. 
Thus, in this work, we consider the problem of distributing quantum circuits across a quantum network with the optimization objective of minimizing the circuit execution time, which includes latency incurred in generating the required entanglements to execute the remote gates under decoherence constraints.

Distributing quantum circuits entails mapping the circuit's qubits to the quantum network's qubit memories and introducing quantum communication operations to execute remote gates 
(gates spanning multiple QCs). Thus, in our proposed approaches, we first determine
an efficient allocation of qubits to memories in the computers such that the estimated
estimated execution time is minimized, and after having established the mapping of qubits, 
we use efficient strategies to execute the gates using required entanglements with minimum 
latency under network resource and decoherence constraints.
The allocation of qubits to memories also entails mapping the qubits within each computer since limited connectivity within computers can affect the circuit execution time.

\para{Prior Work.}
The problem of distributing
quantum circuits in quantum networks has gained significant attention in recent years, resulting in the development
of efficient solutions tailored to various settings and objectives. However, almost all works on distributing  quantum circuits have focused on the objective of minimizing the number of 
maximally-entangled pairs (\EPs) either by minimizing the number of 
cat-entanglements~\cite{Andres-MartinezH:19, g2021efficient} or the number of teleportations~\cite{zomorodi2018optimizing, sundaram2023distributing, Nikahd_2021}.
In the work closest to ours, given an allocation of qubits to computers, \cite{Caleffi} minimizes 
the number of time slots to generate EPs required to execute the remote gates; they make the simplistic assumption that 
each EP’s generation takes a single unit of time.
In our work, we aim to comprehensively solve the problem of the distribution of quantum circuits by considering the optimization objective of minimizing the circuit execution time under the network resource and decoherence constraints. 
{\em The circuit execution time must include generation latencies of the required EPs, which must take into consideration the stochasticity of the underlying processes; this makes the problem particularly challenging and significantly different from prior works on distributing quantum circuits.}
When allocating the qubits to network memories, we also 
map qubits to memories {\em within each computer} as limited connectivity 
within computers also affects the circuit execution time.


\para{Our Contributions.}
In this paper, we formulate the problem of distributing quantum circuits in quantum networks to minimize circuit execution time under given constraints. We address this problem in two
steps, as below.

\begin{itemize}
    \item For the first step of allocating circuit qubits \blue{to} network memories, we develop a heuristic scheme based on an approximation algorithm for a special case of the well-known maximum quadratic assignment problem. (\S\ref{sec:step1}).
    
    \item For the second step of developing an efficient execution scheme for a given qubit allocation, we design the following algorithms. 
    \begin{itemize}
        \item 
        For the special case, when the consumption order of the required EPs is total, we develop a provably optimal dynamic programming approach; we generalize it to the general case.
        
        \item 
        For the special case, when the consumption order of the required EPs is null, we develop a greedy heuristic with appropriate performance guarantees under reasonable assumptions; we generalize it to the general case.
        
        \item 
        When using cat-entanglements to execute remote gates, we develop a scheme for selecting a provably near-optimal set of cat-entanglements. (\S\ref{sec:ce})
    \end{itemize}
    
    \item Finally,  we demonstrate the effectiveness of our developed techniques by evaluating them on randomly generated circuits and known benchmarks and show that our techniques outperform the prior work by up to $95\%$. (\S\ref{sec:eval})
\end{itemize}

%% file: 2-Background.tex
\section{\bf Background}
\label{sec:back}

\para{Quantum Circuit Representation.}
We represent an \emph{abstract quantum circuit} $C$ over a set of qubits $\mathcal{Q} = \{q_1, q_2, \ldots\}$ as a sequence of gates $\langle g_1, g_2, \ldots \rangle$ where each $g_t$ is either binary \CNOT gate or a unary gate (a universal set of gates).
We represent binary gates as triplets $(q_i, q_j, t)$ where $q_i$ and $q_j$ are the two operands and $t$ is the time instant of the gate in the circuit, and unary gates as pairs $(q_i, t)$ where $q_i$ is the operand and $t$ is the time instant. 

\para{Quantum  Network (QN).}
A quantum network is a network of quantum computers (QCs) represented as a connected graph with nodes as QCs and edges representing (quantum and classical) communication links. Each computer has a certain amount of quantum memory to store the data/circuit qubits.

\subsection{\bf Distribution of Quantum Circuits over QNs}
\label{sec:dqc}

Given a quantum network (QN) and a quantum circuit, \hide{our goal is}\blue{we seek} to 
distribute and execute the given circuit over the network in a way that the
execution incurs minimum time. 
Distribution of a circuit
over a network essentially entails two aspects: (i) distributing the
circuit qubits across the QCs/nodes of the QN, and (ii) executing the
``remote'' binary gates (i.e., gates with operands on different QCs) efficiently. 

\para{Executing Remote Quantum Gates.}
To execute such remote gates, we need to bring all operands' values into a single QC via quantum communication. 
Since direct transmission of qubit
is subject to unrecoverable errors,
we consider the following ways to 
communicate qubits across network nodes.

\begin{figure*}
\vspace{-0.35in}
  \begin{minipage}[t]{.67\linewidth}
    \includegraphics[width=\linewidth]{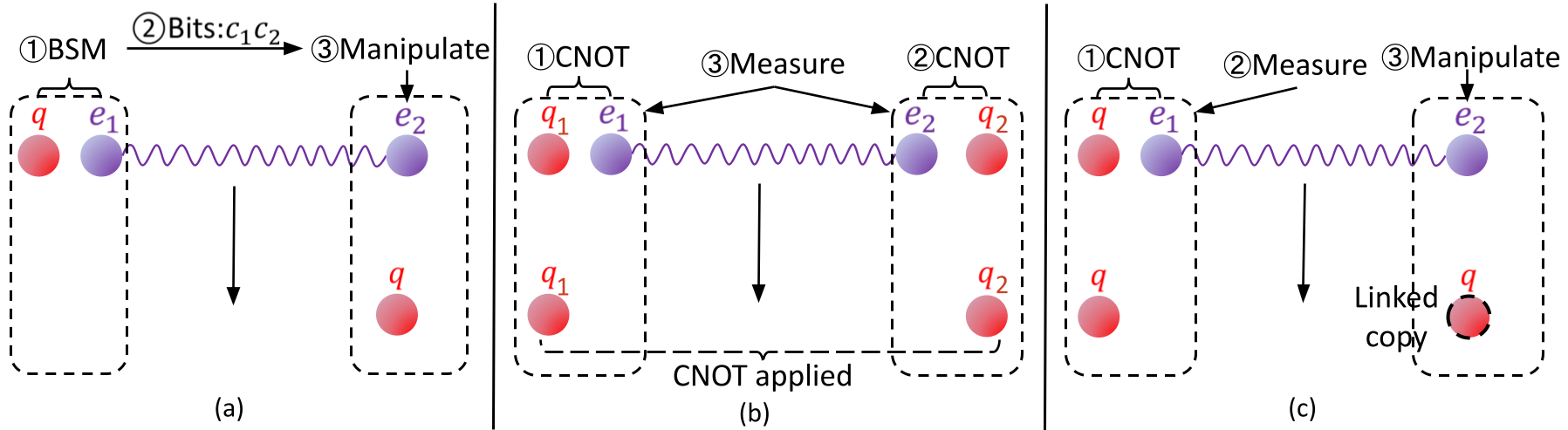}%
    \caption{Quantum communication. (a) Teleportation, (b) Telegate, (c) Cat-Entanglement. Dashed boxes are the network nodes; the initial (final) state of the qubits is at the top (bottom).}%
    \label{fig:Communication}
  \end{minipage}\hfil
  \begin{minipage}[t]{.32\linewidth}
    \includegraphics[width=\textwidth]{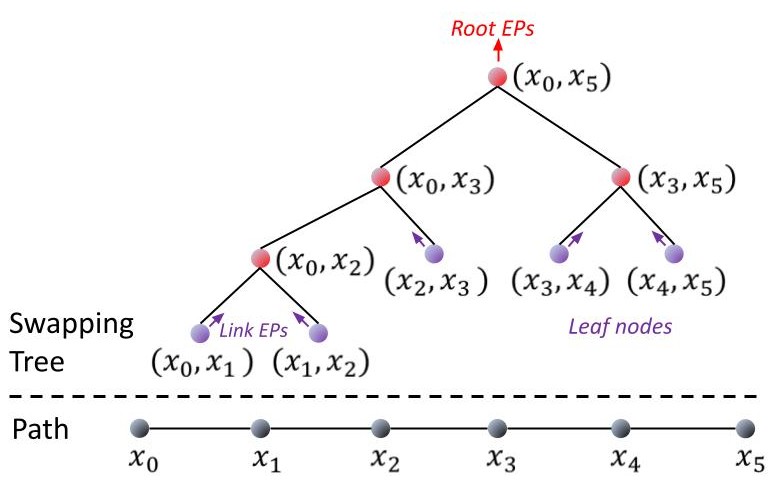}
    \caption{Swapping tree to generate remote EPs.}
    \label{fig:swapping_tree}
  \end{minipage}%
  \vspace{-0.2in}
\end{figure*}

\softpara{Teleportation.}
An alternative approach to physically transmit a qubit from a node
$A$ to $B$ is via \emph{teleportation}~\cite{Bennett+:93} 
which requires an a priori distribution of an \EP over $A$ and $B$. 
See Fig.~\ref{fig:Communication}(a).
\eat{With an \EP distributed over $A$ and $B$, teleportation of a qubit 
state from $A$ to $B$ can be accomplished using classical 
communication and local gate operations while consuming/destroying 
the \EP. }
Teleportation can be used to bring operands of a remote gate
to a single QC for local execution. 

\softpara{Telegates.} 
One way to execute remote gates without communicating the
gate operands are via the telegate protocol~\cite{Caleffi}. 
The telegate protocol (see Fig.~\ref{fig:Communication}(b))
is a sequence of local operations that effectuates the execution of
a remote \CNOT gate with operands at nodes $A$ and $B$; the protocol
requires an a priori distribution of an \EP over $A$ and $B$.

\softpara{Cat-Entanglement: Creating ``Linked Copies'' of a Qubit.}
Yet another approach to execute remote gates is by creating \emph{linked read-only copies} of a qubit across QCs via \emph{cat-entanglement} operations~\cite{Eisert+:00,YimsiriwattanaL:05}. Creating a linked copy of a qubit $q$ at node $A$ to a node $B$
requires a priori distribution of an \EP over $A$ and $B$.
See Fig.~\ref{fig:Communication}(c).
The linked copy at $B$ can be used to execute binary gates 
where $q$ is the (read-only) control qubit, 
until a unary operation needs to be executed on $q$. 
When a unary operation needs to be executed on $q$, a disentanglement
operation is done, which destroys the linked copies, and then, the unary
operation can be done on the original qubit $q$ at $A$.

\softpara{Teleportation vs.\ Telegates vs.\ Cat-Entanglements.}
First, observe that each of these three communication protocols requires a single
EP. Teleportation transports the qubit operand to the computer where the
gate is executed. In contrast, the telegate protocol executes a remote gate
without moving qubits. The main advantage of cat-entanglement is that one cat-entanglement (and, thus, one linked copy) can sometimes be used to execute
several binary gates; however, 
cat-entanglements are best 
used for circuits with only \cz binary gates (see \S\ref{sec:ce}).
This paper considers both telegates and cat-entanglements (\S\ref{sec:ce}) 
for executing remote gates. The use of teleportations leads to dynamic qubit 
allocation (see \S\ref{sec:problem}), which requires more a sophisticated problem 
formulation
and solution, and is thus deferred to our future work.

\subsection{\bf Generating \EPs over Remote Nodes}
\label{sec:eps}


Each of the above communication modes requires an apriori distributed EP. One simple way to have an \EP distributed over nodes $A$ and $B$ is to generate an \EP locally (at some node) and transport the qubits to $A$ and $B$ respectively; 
however, this involves direct communication of qubits, which may not be feasible due to irrecoverable errors over large distances.
To circumvent this challenge, to distribute an \EP over remote nodes, we generate \EPs over larger and larger distances from shorter-distance \EPs using a sequence of ``entanglement-swapping'' operations. 
An entanglement swapping (\es) operation can be looked up as being performed over three nodes $(A, B, C)$ with two \EPs over $(A, B)$ and $(B,C)$; the 
\es operation results in an \EP over the nodes $(A, C)$, by essentially teleporting the first qubit in node $B$ (i.e., the qubit of the EP over $(A,B$)) to the node $C$ using the second \EP over $(B,C)$.
In general, an \EP over a pair of remote nodes $A$ and $B$ can be generated by
a sequence of \es operations over the \EPs over adjacent 
nodes along a path from $A$ to $B$.
For example, see Fig.~\ref{fig:swapping_tree} which shows a ``swapping'' tree
to generate an EP over $(x_0, x_5)$ using EPs along the path $x_0, x_1, x_2, x_3, x_4, x_5$.
Different swapping trees over the same path $P$ can incur different generation
latencies~\cite{tqe-22}.


\softpara{Generation Latency of \EPs.}
The stochastic nature of the \es operations
means that an \EP at the swapping tree's root is successfully 
generated only after many failed attempts; thus the generation
of an EP incurs significant \textit{generation latency}.
To generate a set of \EPs concurrently, we need to allocate the network 
resources appropriately to each \EP's generation. 


\subsection{\bf Decoherence and Fidelity}

\emph{Fidelity} is a measure of how close a realized state is to the ideal. The fidelity of a quantum state decreases over time due to interaction with the environment and gate operations. Time-driven fidelity degradation is called \emph{decoherence}. To bound decoherence, we limit the aggregate time a qubit spends in a quantum memory before being consumed by the {\em decoherence threshold} of $\tau$ seconds.
We address fidelity degradation by using entanglement purification~\cite{Briegel1998QuantumRT} techniques which entails using multiple copies of an EPs to improve its fidelity.

%% file: 3-Problem.tex
\section{\bf Problem Formulation and Related Work}
\label{sec:problem}

We start by defining some terms and models before moving on to the problem formulation.

\para{Qubit-Allocation Function.}
Let the set of network nodes/QCs in the given quantum network be $\mathbb{P} = \{P_1, P_2, \dots, P_N\}$, where each QC $P_i$ is equipped with a set $Q(P_i)$ of qubit memories for data storage.
For a given quantum circuit $C$, let $Q(C)$ be the set of qubits in $C$. 
To execute a circuit over a quantum network, we must first distribute the circuit qubits over the qubit memories in the network nodes. 
To that end, we define the {\em qubit allocation function} $\eta$ as the  function that maps the circuit qubits $Q(C)$ to the network qubits $\bigcup_i Q(P_i)$, i.e., $\eta: Q(C) \rightarrow \bigcup_i Q(P_i)$; the qubit-allocation function must be one-to-one, i.e., only one circuit-qubit can be mapped to a qubit memory in the network.

\softpara{Remote and Local Gates.}
For a given circuit $C$ and a qubit allocation function $\eta$, a gate $(q_i, q_j, t)$ in $C$ is considered to be {\em remote} if $q_i$ and $q_j$ are assigned to different {\em computers} by the qubit allocation function, i.e., $\eta(q_i) \in Q(P_i)$ and $ \eta(q_j) \in Q(P_j)$ where $i \neq j$. 
Every gate that is \emph{not} a remote gate is considered a \emph{local gate}.

\para{Coupling Graph; Executing Local Gates Using Swaps.} 
In realistic quantum computer hardware, quantum gates can be executed only
over those qubit operands that are in memories located ``close by.'' 
To characterize this limitation, we use a concept of {\em coupling graph} $U_i$ 
over the set of qubits, $Q(P_i)$ for each computer $P_i$; an edge $(m_a, m_b)$ in $U_i$ signifies that the memories $m_a$ and $m_b$ are located close by that qubits in them can be operated upon directly using a binary gate.
If two qubits $q_c$ and $q_d$ are in far away memories $m_c$ and $m_d$ within a computer $P_i$, then to execute a gate $(q_c, q_d, t)$, we need to move the qubits close-by through a series of swap operations. In particular, we assume that the swaps are done along the shortest path connecting $m_c$ and $m_d$, and thus, number of swaps needed is equal to the shortest path length between $m_c$ and $m_d$ in the coupling graph $U_i$. 

\para{Enforcing Static Qubit-Allocation Function.} In general, the qubit-allocation function may change during the circuit execution due to teleportations and/or swap operations. However, to simplify algorithm design and analysis, we enforce the qubit allocation to be fixed throughout the circuit execution; we relax this enforcement 
in our implementation and evaluations (\S\ref{sec:eval}). 
This enforcement of static qubit allocations entails the following:
\begin{enumerate}
\item
We don't need to consider teleportations to execute remote gates, as they would incur double the cost than telegates, as the qubits will need to be teleported back and forth. Thus, to execute remote gates, we use telegates or cat-entanglements (\S\ref{sec:ce}). 

\item
For the execution of local gates that require swaps, we reverse the sequence of swap operations after the gate execution to preserve the qubit allocation function. We relax this in our evaluations. 
\end{enumerate}
We now formulate the problem of distributing quantum circuits in quantum networks with minimum circuit execution time under given constraints.

\para{Distributed Quantum Computation and Routing (\dqcqr) Problem.} 
Given a quantum network and a quantum circuit $C$, the \dqcqr problem is to distribute and execute $C$ over the given network with minimum execution time under the network constraints (in particular, memory at each network node, EP generation resources, and decoherence). The \dqcqr problem 
entails (i) finding a qubit-allocation function, which remains fixed throughout the circuit execution (see above), and (ii) an execution scheme involving the execution of local and remote gates, such that the total circuit execution time is minimized. As discussed above, executing local gates may involve swaps, while executing remote gates requires communication, which requires \EPs to be generated, as discussed below.

The \dqcqr problem can be shown to be NP-Hard by a reduction from the min quadratic-assignment problem~\cite{QAPHardness}.

\softpara{Key Challenges: Qubit-Allocation, Generation Order of EPs.} The key initial task of the \dqcqr problem is to determine the qubit-allocation function that facilitates a minimum-time execution scheme. Now, given a qubit-allocation function, the resulting set of remote gates determines the \EPs that need to be generated. Without decoherence (and minimal/zero gate and swap latencies), the desired \EPs can be generated at the start of a circuit execution in any order. 
However, {\bf {\em in the presence of decoherence, the required \EPs need to be generated in an appropriate order}}---so that each generated \EP can be consumed before decohering.\footnote{Note that the order of gates in the given circuit imposes a consumption order on the \EPs which, in turn, imposes a generation order on \EPs and may also require them to wait before being consumed.}
In the case of non-zero gate execution and swap latencies, generation of \EPs can also be overlapped with swap and gate operations to minimize overall execution time. 
In summary, the key challenges in solving the \dqcqr problem are to determine an efficient qubit-allocation function and an execution scheme; the latter largely involves determining 
when and how to generate the required \EPs.

\begin{figure*}[t]
\vspace{-0.2in}
\centering
\includegraphics[width=0.9\textwidth]{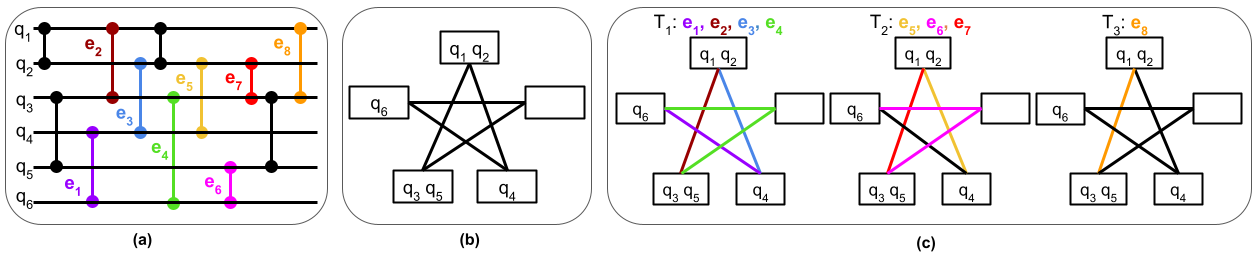}
\vspace{-0.1in}
\caption{\dqcqr Problem Example~1. {\bf (a)} Quantum circuit, with the remote gates (based on the qubit allocation in (b)) colored and labeled with the \EPs required. {\bf (b)} Qubit allocation over the quantum network. {\bf (c)} Execution scheme, i.e., the batches and order in which the required \EPs are generated; the colored paths in (c) represent the entanglement routes along which the corresponding \EPs are generated.}
\vspace*{-0.2in}
\label{fig:DQC_QRExample}
\end{figure*}

\para{Example 1.} Fig.~\ref{fig:DQC_QRExample} shows an instance of the 
\dqcqr problem, with the input circuit of five
qubits in Fig.~\ref{fig:DQC_QRExample} (a). 
The colored gates (non-black) represent remote gates based on the qubit allocation over a quantum network shown in Fig.~\ref{fig:DQC_QRExample}(b).
Fig.~\ref{fig:DQC_QRExample}(b) doesn't show the coupling graph within each network node, for the sake of clarity, as each node only holds at most two qubits. 
Fig.~\ref{fig:DQC_QRExample} (c) shows one possible execution scheme for the given qubit allocation: Execute the EPs in three batches in order: $\{e_1, e_2, e_3, e_4\}, \{e_5, e_6, e_7\}, \{e_8\};$ within each batch, the EPs can be generated concurrently. As EPs get generated, they are consumed to execute the corresponding remote gates.


\subsection{\bf Related Work}
\label{sec:related}

The \dqcqr problem is related to 
two studied problems, 
viz., Distributing Quantum Circuits and 
the Qubit Routing. We discuss these below,
then discuss the work closest to ours.

\softpara{Prior Work on Distributing Quantum Circuits (\dqc) 
Problem.} The \dqc problem is 
to distribute a given quantum circuit over a quantum network with some 
optimization objective. In contrast to our \dqcqr problem, the \dqc 
problem ignores the coupling graphs within each 
computer~\cite{Andres-MartinezH:19, g2021efficient,Daei2020OptimizedQC,Nikahd_2021}, and thus, 
the swap operations needed. The optimization objective used in almost all 
of these works has been to minimize the communication 
cost---defined as the {\em number} of EPs used; a couple of recent works~\cite{MaoProbability,MaoQubitAllo} consider EPs with arbitrary (but fixed) cost and 
develop a simulated-annealing heuristic for the qubit-allocation part of the 
\dqc problem.
Our schemes in this work use a similar two-step strategy as some~\cite{g2021efficient,dqc-gen,sundaram2023distributing} of the \dqc works, but otherwise differ significantly as we need to consider  the stochastic and concurrent generation of required EPs to minimize execution time, and the coupling graph; we also consider telegates to execute remote gates while the two-step \dqc works consider entanglements~\cite{g2021efficient,dqc-gen} and/or teleportations~\cite{dqc-gen,sundaram2023distributing}.

\softpara{Prior Work on Qubit Routing (\qr) Problem.} The \qr problem considers limited connectivity across memories within a single QC and entails finding an allocation of circuit qubits to qubit memories,
followed by a minimal sequence of swap operations to execute the circuit gates. 
Most \qr works~\cite{tket, Zulehner, Sabre} first find an initial allocation of circuit qubits to memories and then add
swap gates as needed. The optimization objective is to minimize the number of swap operations or the circuit depth overhead. 
Some works~\cite{Siraichi, Childs, Cowtan2019} have used subgraph isomorphism to obtain high-quality qubit allocations followed by token-swapping heuristics to determine a near-optimal set of swap operations.  
Our work generalizes the \qr problem by considering a {\em network}
of QCs, and thus, needs to address the generation of EPs as well as swap 
operations. We note that our treatment of qubit routing is limited as
we keep the qubit allocation fixed (this is relaxed in our evaluations (\S\ref{sec:eval}).

\softpara{Prior Work on Our \dqcqr Problem.} 
The works closest to ours~\cite{Caleffi, Ferrari-etal-2021} address the
\dqcqr problem with some limitations and differences. In particular,~\cite{Ferrari-etal-2021} focuses on estimating 
only the {\em worst-case} overhead in executing a circuit over a network,
by the worst-case linear network topology; the overhead considered is in terms of an increase in circuit ``layers'' and EPs required to execute remote gates.
In the closest work that inspires this work,~\cite{Caleffi} focuses on the efficient execution of remote gates using telegates, {\em given} a qubit allocation (they consider qubit allocation to be out of the scope of their work), to minimize the number of circuit layers; they assume generation latency of each EP to be uniform (one time slot), and ignore decoherence constraints.  In \S\ref{sec:eval}, we compare our approaches with their approach (supplemented with our qubit-allocation strategy).
The main difference between our work and~\cite{Caleffi} is that we consider qubit-allocation and the stochastic generation of EPs under decoherence and network resource constraints. We also consider cat-entanglements for executing remote gates, which result in significant 
performance improvement, as demonstrated in our evaluations.

%% file: 5-Step-1.tex
\section{\bf High-Level Approach}
\label{sec:high-level}

In the previous section, we formulated the \dqcqr problem as distributing a given circuit over a given network with minimum execution time. As mentioned above, the key challenges or subproblems in solving the \dqcqr problem is to determine the qubit-allocation function and the execution scheme such that the execution time is minimized. 
Thus, it is natural to tackle the \dqcqr problem as a sequence of two steps, similar to the approaches taken in prior works on the similar problem of distributing quantum circuits~\cite{g2021efficient,dqc-gen,sundaram2023distributing}. 

\begin{figure}
\vspace{-0.2in}
\hspace{-1.5in}
\includegraphics[width=\textwidth]{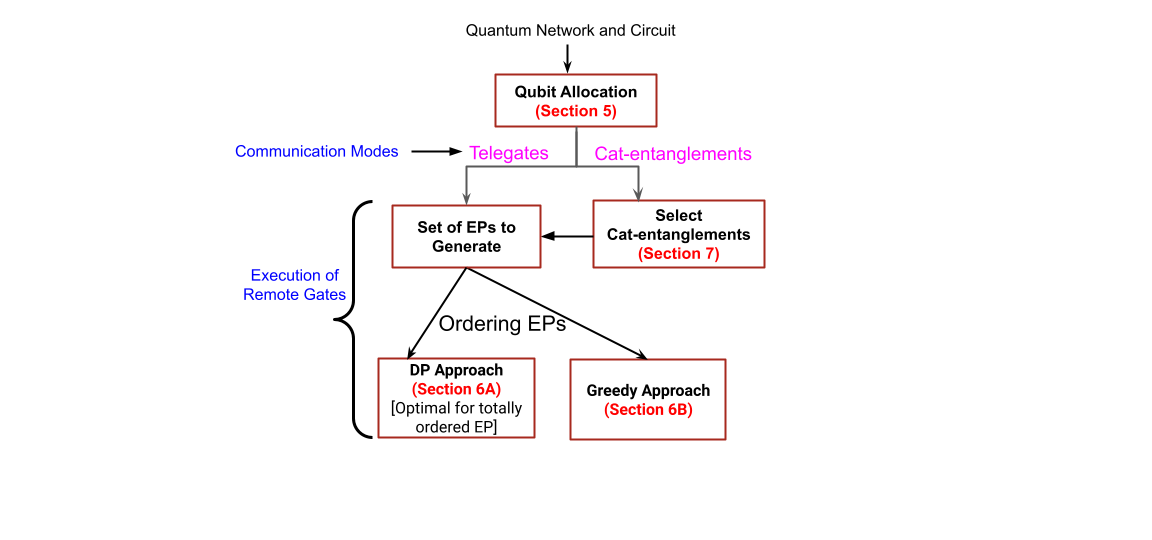}
\vspace{-0.8in}
\caption{Overall plan and organization of the schemes developed.}
\vspace{-0.2in}
\label{fig:plan}
\end{figure}

\begin{enumerate}
\item 
{\em Determine the Qubit-Allocation Function.} 
In this step, given the quantum circuit and the quantum network, we determine the qubit-allocation function that will yield an execution scheme in the second step with minimum total circuit execution time. 

\item 
{\em Determine the Execution Scheme, Given the Qubit Allocation.} 
In this step, given a qubit-allocation function for a given circuit, we determine the execution scheme---which largely entails determining when and how to generate the \EPs required to execute the remote gates. Local gates are executed using appropriate swap operations. 
\end{enumerate}
We discuss the above steps in the following sections. See Fig.~\ref{fig:plan} for the overall plan and organization of the following sections.

\vspace{-0.2in}
\section{\bf Step 1: Determining Qubit Allocation Function}
\label{sec:step1}

In this section, we design an algorithm for determining a
qubit-allocation function to yield an efficient execution scheme. We start with formally defining the qubit-allocation problem and show that it is NP-Hard to approximate
within a constant factor. Then, we design an algorithm for
qubit allocation based on a 4-factor approximation 
algorithm for a special case of the related quadratic-assignment problem. 

\para{Qubit-Allocation Problem Formulation.}
Informally, the qubit-allocation problem essentially maps a circuit graph (defined over circuit qubits, with edge weights representing the number of gates between the qubit-pairs) onto a network-coupling graph (defined over network memories, with edge weights representing the cost of executing a gate over the qubits in the corresponding network memories); the optimization objective is to minimize an appropriately defined cost of the mapping. In effect, we want to map the circuit qubits to network memories so that qubit-pairs with more gates between them are mapped to memories that are ``close by''\hide{ (e.g., close-by in the same computer)}; this can be formally captured by defining a mapping cost
as the weighted sum of the product of the weights of the edge pairs that map to each other in the mapping. We formalize the above intuition below by defining the input graphs and the problem.


\softpara{Circuit Graph $G_C$.} The circuit graph is an edge-weighted graph defined over the set of qubits $Q(C)$ of a given circuit $C$ with weight $w(q_i, q_j)$ associated with an edge $(q_i, q_j)$ representing the number of (binary) gates 
in 
$C$ over the qubits $q_i$ and $q_i$. 

\softpara{Network-Coupling Graph $G_N$.} The network-coupling graph is an edge-weighted graph defined over the set of qubit-memories 
$\bigcup_i Q(P_i)$ in the given quantum network with QCs $\{P_i\}$.
The weight $w'(m_a, m_b)$ 
associated with an edge between two memories $(m_a, m_b)$ is defined as follows.
\begin{itemize}
    \item If $m_a$ and $m_b$ are in the same computer $P_i$, then the weight is the product of swap-operation latency and twice the shortest distance between $m_a$ and $m_b$ in the coupling graph $U_i$ of $P_i$; note that this represents the time to execute the gate over qubits in $m_a$ and $m_b$ using swap operations.
    \item If $m_a \in Q(P_i)$ and $m_b \in Q(P_j)$, then the weight on the edge $(m_a, m_b)$ is the latency of (independently) generating an \EP over network nodes $P_i$ and $P_j$.
\end{itemize}
For the sake of simplicity, the above weighting of the network-coupling graph assumes that each EP is generated independently. However, when we actually generate the EPs in the following step (Step~2 discussed in~\S\ref{sec:step2}), we generate 
them as concurrently as possible. We now formally define the qubit-allocation 
problem.

\softpara{Qubit-Allocation Problem Formulation.} 
Given a quantum circuit $C$ and a quantum network, the qubit-allocation problem is to determine a mapping $\eta: Q(C) \rightarrow \bigcup_i Q(P_i)$ from the circuit qubits to the qubit-memories of the given network, such that the mapping-cost ${\rm cost}(\eta)$, defined as below, is minimized.
\begin{center}
    ${\rm cost}(\eta) = \sum_{q_i, q_j \in Q(C)} w(q_i, q_j) \times w'(\eta(q_i), \eta(q_j))$
\end{center}
The qubit-allocation problem can be shown to be NP-Hard by a reduction from the well-known {\tt minimum quadratic assignment} problem (\minqap) ~\cite{QAPHardness}. In addition, by the same reduction, it can also be shown that there is no constant-factor polynomial-time approximation algorithm for the qubit-allocation problem unless P=NP. 

Below, we design an algorithm to solve the qubit-allocation problem. We start with a formal definition of the min-QAP problem since our designed algorithms are based on it.

{\em Minimum Quadratic Assignment Problem.} Given two $n \times n$ matrices $A$ and $B$, the
\minqap problem is to permute the rows/columns in $B$ such that the sum of the product of the matrix elements is minimized. More formally, the goal is to find a permutation (a one-to-one mapping) $P: \{1, 2, \ldots, n\} \mapsto \{1, 2, \ldots, n\}$ to minimize the following quantity: $\sum_{1 \leq i,j \leq n} A[i,j] B[P(i), P(j)]$.

\para{Qubit-Allocation Algorithm Based on Quadratic-Assignment.}
The qubit-allocation problem is quite similar to the minimum quadratic assignment problem (\minqap) defined above, except that the \minqap problem requires the two input matrices/graphs to be complete and of equal sizes. Moreover, the \minqap problem is known to be inapproximable within any polynomial factor~\cite{QAPHardness} (unless P = NP), but the {\em maximum} quadratic assignment problem (\maxqap) has a 4-factor approximation algorithm when one of the graphs satisfies the triangular inequality~\cite{Esther}. Based on the above, we design an efficient heuristic for our qubit-allocation problem as follows.

Let $C$ be the given quantum circuit with $G_C$ as the corresponding circuit graph. 
Let $G_N$ be the network-coupling graph of the given quantum network. Without loss of generality, let us assume that the size of $G_C$ (i.e., the number of circuit-qubits $|Q(C)|$) is less than the number of qubit-memories in the network. Our qubit-allocation algorithm is:
\begin{enumerate}[(a)]
\item 
Add dummy nodes and edges to $G_C$ to create a new graph $G'_C,$ which is the same size as the network-coupling graph $G_N$. In particular, we add an appropriate number of dummy nodes to $G_C$ and connect these nodes with edges of zero weight to the original nodes in $G_C$. We also add edges of weight zero between all pairs of dummy nodes.

\item 
Create a new graph ${G''_C}$ by replacing each edge-weight $w(q_i, q_j)$ with $M-w(q_i,q_j)$, where $M$ is a sufficiently large number (e.g., equal to the total sum of edge-weights in $G_C$). This change of weights converts the minimizing mapping-cost problem in $G'_C$ to the maximizing mapping-cost problem in $G''_C$.

\item 
Solve the \maxqap over the input graph $G''_C$ and $G_N$ using the 4-approximation algorithm for \maxqap. 
\end{enumerate}
\cblue
\input{5-step-triangle}
\cblack



%% file: 5-step-triangle.tex
\begin{thm-w-prf}
The 4-approximation \maxqap algorithm from~\cite{Esther} returns a $6/p$-approximate solution for the \maxqap problem over $G''_C$ and $G_N$; here,
$p$ is the entanglement-swapping's success probability.
\label{thm:ce}
\end{thm-w-prf}
\begin{prf}
    {\em (Sketch)} First, we show that the edge weights in $G_N$ satisfy the triangular inequality within a constant factor. If $m_i$ and $m_j$ are  memories in a computer $P$,
then $w'(m_i,m_j)\leq w'(m_i,m_k)+w'(m_k,m_j)$ for any memory $m_k$ in $P$ since 
$w'(m_i,m_j)$ is the time to perform swaps along the shortest path between $m_i$ and $m_j$. If $m_i$ and $m_j$ 
are in different computers $P_1$ and $P_2$, then, for any $m_k\in P_3$, we 
have~\cite{tqe-22}: 
$$w'(m_i,m_j)\leq 3/(2p)(w'(m_i,m_k)+w'(m_k,m_j)),$$ since
an EP over $(P_1, P_2)$ can be generated using EPs over $(P_2, P_3)$ and $(P_1, P_3)$.
The $3/(2p)$ factor above, when incorporated in the analysis for the 4-approximation result from~\cite{Esther}, yields the $4(3/(2p))=(6/p)$ approximation factor. 
\end{prf}

We note that our overall qubit-allocation algorithm (i.e., (a)-(c) above) is still a heuristic since the 
conversion from the \minqap to \maxqap does not preserve the approximation factor; however, the algorithm performed well empirically (\S\ref{sec:eval}).



%% file: 6-Step-2.tex
\section{\bf Step 2: Executing Remote Gates by Generating \EPs \eat{Concurrently}}
\label{sec:step2}

Step~1, discussed in the previous section, determines the qubit allocation function, which, in turn, determines the set of remote gates that need to be executed using quantum communication mechanisms. In this section, we use telegates to execute remote gates (we consider cat-entanglements in  \S\ref{sec:ce}); since telegates require an a priori \EP distributed over the nodes/QCs that contain the gate operands---for each remote gate, we need to generate an EP over the pair of nodes that contain the 
gate operands. 
For simplicity, we assume the swap and gate execution latencies to be zero 
(i.e., negligible compared to the EP generation latencies); we relax this assumption 
in \S\ref{sec:gen}.
Under the above assumption, given a qubit allocation, the problem of executing the given 
circuit with minimum execution time essentially boils down to generating (possibly, concurrently) the required EPs in minimum total latency 
(under the constraints of consumption order and decoherence thresholds, as discussed below).
We start by considering the simpler case when there is no decoherence.


\para{Generating EPs When No Decoherence (\ges).} 
Let us use \E to denote the set of \EPs required to be generated for a given
qubit-allocation function. 
Without decoherence constraints, Step 2 of our approach boils down to generating the  \EPs in \E with minimum total latency. 
This is because if all the \EPs in \E  can be generated in total time $T$, then the total circuit can also be executed in $T$ as all the circuit gates can be executed instantly right after all the \EPs are generated; the converse is also true, i.e., if the circuit execution time is $T$, then it implies that
 the total generation latency of all the \EPs in \E is less than $T$.
To generate a set of \EPs \E concurrently with minimal total latency without decoherence, 
we use the linear programming (LP) approach from~\cite{DPpaper} 
that generates a set of desired EPs using swapping trees with a maximum total generation rate. 
In our context, since we are interested in only generating a {\em specific set} of EPs with a minimum 
total {\em latency}, we modify the LP appropriately after each EP has been generated successfully.
We refer to the above procedure to generate a set of EPs concurrently as \ges.

\begin{figure}[t]
\vspace*{-0.4in}
\centering
\includegraphics[width=0.4\textwidth]{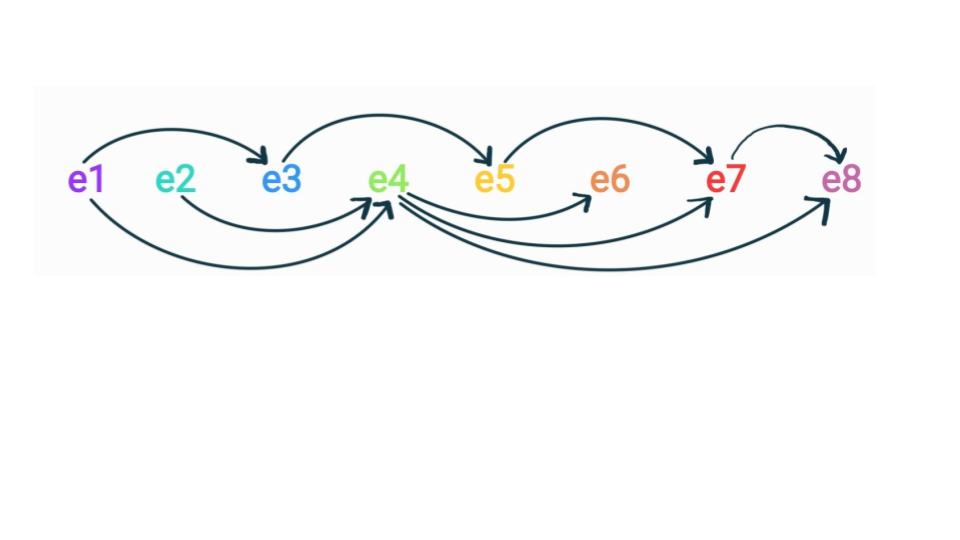}
\vspace*{-2cm}
\caption{Consumption order of \EPs required in Example 1. A directed edge from $e_i$ to $e_j$ indicates $e_i$ must be consumed before $e_j$.}
\label{fig:ep-order}
\vspace*{-0.28in}
\end{figure}

\para{Generating EPs under Decoherence Constraints.}
If there is decoherence, EPs can decohere while and after (as they wait to be consumed) being generated. In particular, decoherence may require that we partition \E into subsets of EPs (called {\em batches}) with each batch generated independently---as
generation of the entire set \E concurrently may result in some EPs
having a generation latency of more than the decoherence threshold $\tau$.
Once the set \E has been partitioned appropriately into batches, the batches need to be generated (with each batch generated independently) in an appropriate order to prevent EPs from decohering. This is because if we generate the batches in an arbitrary order, then some already-generated EPs may have to wait for other EPs to be generated and consumed before being consumed, leading to the waiting EPs possibly getting decohered. 
For example, for the example in Fig.~\ref{fig:DQC_QRExample}, consider two batches of EPs:
$\E_1 = \{e_1, e_2, e_5\}$ and $\E_2 = \{e_3, e_4, e_6\}$. Based on the consumption order (formally defined below) in
Fig.~\ref{fig:ep-order}, if $\E_1$ is generated before $\E_2$ then $e_5 \in \E_1$ will need to wait
for $e_3 \in \E_2$ to be generated and consumed, before $e_5$ can be consumed. In this case, even
generating $\E_2$ before $\E_1$ will lead to an EP waiting for another.
In general, partitioning \E into batches and the order in which these batches are generated depends on the consumption order constraints between the EPs, which comes from the order of the corresponding gates in the circuit. Below, we formally define the consumption order over \E, problem formulation of generating \E under given constraints, and then present our approaches.

\softpara{Consumption Order \C\ over EPs.}
Consider the set of \EPs \E that must be generated to execute the remote gates for a given qubit allocation. The consumption order \C over the \EPs in \E is defined as follows. For $E_i, E_j \in \E$, we have $E_i \prec E_j$ if the gate $g_i$ and $g_j$ in $C$ corresponding to the \EPs  $E_i$ and $E_j$ are such that they share an operand and $g_i$ occurs before $g_j$ in $C$. In addition, \C is closed under transitive closure, i.e., if $E_i \prec E_j$ and $E_j \prec E_k$, then $E_i \prec E_k$. Note that the order \C is a strict partial order; in particular, for no $E_i$, $E_i \prec E_i$.

\para{Generation of EPs under Decoherence (\ged) Problem.}
Given a set \E of EPs with a consumption order \C, the EG problem is to partition the set  \E into subsets/batches $\E_1, \E_2, \ldots, \E_l$ (generated in that order) such that 
(i) each subset $\E_i$ has a 
total generation latency of at most $\tau$, and (ii) the total circuit execution 
latency is minimized, assuming the batches are generated in the order $\E_1, \E_2, \ldots, \E_l$ and we use the \ges procedure to generate each batch.

\softpara{No-Wait Property of \ged Solutions.} A \ged solution $\E_1, \E_2, \ldots, \E_l$ satisfies the no-wait property if for all $E_i, E_j$ pairs of EPs in \E, if $E_i \prec E_j$, then $E_i \in \E_i$ and $E_j \in \E_j$ such that $i \leq j$. For example, 
for Fig.~\ref{fig:ep-order}, the \ged solution $\{e_1, e_2, e_3, e_4\}, \{e_5, e_6, e_7\}, \{e_8\}$ satisfies the no-wait property.
\ged solutions with the above no-wait property ensure that the circuit execution time equals the sum of the generation latencies of the batches. Note that, in general, if the no-wait property is not satisfied, then the total circuit execution latency can be more than the sum of the generation latencies of the batches due to some already-generated EPs {\em waiting} for other EPs to be generated. Our designed approaches return solutions with the no-wait property.

\medskip
In the following subsections, we develop dynamic programming (DP) and greedy approaches for the \ged problem. The motivation behind our approaches
is: (i) If the consumption order \C is total, then an optimal approach based on DP can be designed. (ii) If no consumption order exists over EPs, then a greedy approach with performance guarantees under certain reasonable assumptions can be designed. Both approaches can be generalized for general partial consumption orders.

\subsection{\bf Dynamic Programming (DP) for Generating EPs}
\label{sec:DP}

Here, we provide a dynamic programming (DP) algorithm for the \ged problem; 
the DP algorithm is optimal for the special case when the consumption order \C is total. 
We start with presenting the DP algorithm for the special case when the
consumption order \C is total; we then generalize the algorithm for the general \C later.

\para{Optimal DP Algorithm when \C is Total.}
Let the set \E of required \EPs be $\{E_1,E_2,\dots, E_m\}$ and 
without loss of generality, let the 
consumption order \C be $E_1 \prec E_2 \prec E_3 \dots \prec E_m$ and thus, be a total order. 
First, we observe (see Lemma~\ref{lem:dp} below) 
that, if the consumption order \C is total, then 
there is an optimal \ged solution that 
partitions \E into ``contiguous'' batches---i.e., each batch is
of the type $\{E_{i}, E_{i+1}, E_{i+2} \dots E_{j-1}, E_{j}\}$ for $1 \leq i \leq j \leq m$.

\begin{lem-wo-prf} 
\label{lem:dp}
Consider a \ged problem for a set of EPS $\E = \{E_1,E_2,\dots, E_m\}$  
and a total consumption order \C as $E_1 \prec E_2 \prec E_3 \dots \prec E_m$. 
There exists an optimal \ged solution where each batch is a contiguous set of 
EPs.
\end{lem-wo-prf}

Based on the above Lemma, finding the optimal \ged solution for \E with a total order
\C is tantamount to 
partitioning \E into
contiguous batches such that (i) each batch has a generation latency of less than the decoherence threshold $\tau$, and (ii) the sum of generation latencies of the batches is minimized.
Such an optimal partition of \E can be found using a dynamic programming approach as follows. In particular, let us define $S_j$ be the optimal partitioning of the 
subset $\{E_1, E_2, \ldots, E_j\}$ of \EPs into contiguous batches such that 
(i) each batch has a generation latency of less than $\tau$, and (ii) the sum of the latencies of the batches is minimum. 
Let $S[j]$ be the sum of generation latencies of the batches in $S_j$.
Also, let $L_{ij}$ be the total generation latency of the set of EPs 
$\{E_{i}, E_{i+1}, E_{i+2} \dots E_{j-1}, E_{j}\}$ ($1 \leq i \leq j \leq m$) when generated independently using the procedure \ges. Then, we have:
\begin{eqnarray*} 
    S[1] &=&    L_{11} \\
    S[j] &=&    min_{i | i < j\ {\rm and}\ L_{ij} \leq \tau} \ S[i]  + L_{ij}
\label{eqn:simpledp}
\end{eqnarray*}
Using the above recursive equation, we can compute $S[j]$ (and the corresponding partitioning $S_j$) for all $j$, which yields  the optimal solution $S_n$ of the 
\ged problem for $\E$.

\begin{thm-wo-prf}
The above DP algorithm yields an optimal \ged solution when \C is total. 
\end{thm-wo-prf}

\para{DP Algorithm for General \C.} 
To generalize the above DP algorithm for a general consumption order \C, we first create a ``topological sorting'' of the EPs using the ''edges'' in \C and then run the DP algorithm over the sorted/sequenced EPs. More formally, given a set of EPs \E and a consumption order \C over them, we first order and rename the EPs in \E as $\langle E_1, E_2, \ldots, E_m \rangle$ such that for all $E_i, E_j$, if $E_i \prec E_j$ then $i < j$.  Note that such an ordering is always possible since the consumption order \C is a {\em strict} partial order (i.e., the \C relation is acyclic). There are many such orders possible, and we pick any such order. After ordering the EPs as above, we run the above DP algorithm over the ordered list $\langle E_1, E_2, \ldots, E_m \rangle$ to determine the \ged solution. It is easy to see that the DP solution satisfies the no-wait property, even for the general \C. 

\subsection{\bf Greedy Approach for Generating EPs}
\label{sec:Greedy}

We now design a greedy algorithm for the \ged problem. We start by considering
the special case when \C is null. 

\para{\greedy: Greedy Algorithm when \C is Null.}
For the special case when $\C$ is null, at a high-level,
the greedy algorithm iteratively picks a batch of 
EPs (from the remaining EPs in \E) with the lowest average 
latency. It can be shown
that such a greedy algorithm would 
deliver an $O(\log (n))$-approximate solution, where $n$ is the number of gates.
However, determining the batch that has the minimum 
average latency is NP-hard; thus, 
within each iteration, we use an (inner-level) 
greedy approach to determine the batch with {\em near-minimum} 
average latency. We now present our greedy approach (called \greedy) in more detail, and
and present performance guarantee results.
Let $\E=\{E_1,\dots, E_m\}$ be the set of \EPs that need to be generated as an input to the \ged problem, and the consumption order \C be null. Then, our proposed \greedy algorithm is as follows.

\para{\greedy:}
    \begin{enumerate}
        \item Let $S = \E$, the set of remaining \EPs. 
        \item $i=1$.    /* The below generates the batch $\E_i$ */
        \item Initialize $\E_i = E$, where $E$ is the EP in $S$ of lowest latency.
        \item Initialize $S'=S$.
        \item For $k=1$ to $|S|$: /* Iteratively remove an EP $E$ from $S'$ that reduces the latency of remaining $S'$ as much as possible. Update $\E_i$ to $S'$, if needed.*/
        \begin{enumerate}
            \item Remove the EP $E$ from $S'$ that reduces its latency the most. That is, pick an $E$ in $\Bar{S}$ that minimizes Latency$(S' - \{E\})$.
            \item $S'=S'-\{E\}$.
            \item If $S'$ has latency less than $\tau$ and has lower average latency than $\E_i$, then $\E_i = S'$.
        \end{enumerate}
        \item $S = S - \E_i$.
        \item If $S = \{ \} $ RETURN $\E_1, \E_2, \ldots, \E_i$, ELSE Increment $i$ and Go to Step \#3.
    \end{enumerate}

We use \ges procedure to determine the total latency of any set of EPs in the above
\greedy algorithm.
Note that since \C is null, the \greedy solution trivially satisfies the no-wait property; hence, the circuit execution time equals the sum of the latencies of the batches in the solution.

\para{Performance Guarantee Results.} We make the following observations (we omit the proofs here). (i) If, in each iteration above, $\E_i$ is indeed the subset of EPs with the lowest average latency, then the \greedy approach delivers an $O(\log (n))$-approximate solution.
(ii) If the \ges procedure corresponds to a submodular function, then, in each iteration, the selected $\E_i$ is such that, for a given $|\E_i|$, $(S - \E_i)$'s latency is at least 63\% of the maximum possible. In other words, the $\E_i$ is derived
by removing a near-optimal set of EPs from $S$.

\para{\greedyg: Greedy Approach for a General \C.} 
The \greedy algorithm can be generalized to a general consumption order
\C. The key goal of our generalized algorithm (called \greedyg) 
is to preserve the no-wait property of the solution so that the circuit execution time doesn't include 
additional wait times. 
In particular, we modify the above greedy approach as follows. 
In Line~5 of the \greedy's pseudo-code above, which selects near-optimal batch $\E_i$ by iteratively removing
an appropriate EP $E$, we instead remove $E$ with all its ``descendants'' (i.e., all EPs $E_j$'s in $S$ s.t.\ $E \prec E_j$); more formally, in Line~5a, we pick an $E$ that minimizes Latency($S' - E')/|S'-E'|$ where $E'$ is $E$ and all its descendants.
Removal of descendants ensures that the selected batch $\E_i$ doesn't include any EP that ''depends'' on another EP not yet included in a batch---and thus, ensures that the eventual solution satisfies the no-wait property. For example, consider the first iteration of \greedyg over the \EPs in Fig.~\ref{fig:ep-order}. When considering removal of $e_4$, 
we must also remove $\{e_6, e_7,e_8\}$, leaving 
us with $\{e_1,e_2,e_3, e_5\}$. If $e_4$ 
is indeed 
chosen for removal, then we continue the 
algorithm over $\{e_1,e_2,e_3, e_5\}$, while
keeping $\{e_1,e_2,e_3, e_5\}$ as a potential choice for $\E_1$.

%% file: 7a-Cat.tex
\section{\bf Cat-Entanglements to Execute Remote Gates}
\label{sec:ce}

We have considered telegates to execute remote gates until now. We now consider using cat-entanglements instead, which can result in lower circuit execution time since a single cat-entanglement can enable the execution of several remote gates.

Given a qubit allocation, several possible sets of CEs may be sufficient to execute the required remote gates; thus, in this section, we discuss the problem of selecting a near-optimal set of CEs sufficient to execute the remote gates. The selected CEs yield the EPs to be generated; these EPs are then batched and ordered using approaches discussed in the previous section. We start with defining cat-entanglements.


\para{Cat-Entanglements (CEs).} 
A cat-entanglement (CE), requiring one EP, creates a read-only linked copy of a qubit $q$ at another network node. 
Such a linked copy can be used as a control-qubit operand to execute gates until there is a unary gate on the qubit $q$, at which point the linked copy must be ``destroyed'' using the disentanglement operation (which doesn't require an additional EP). 
Thus, though telegates and CEs both require a single EP, a telegate helps
execute one remote gate, while a CE may help execute several remote gates. Thus, using CEs can drastically reduce the set/number of \EPs required and, thus, the circuit execution time.

\softpara{Notation.} A \emph{cat-entanglement} is a triplet $(q_i, P_k, t)$ signifying creation of a linked copy of qubit $q_i$ on
node $P_k$ at circuit's time instant $t$. Disentanglement operations are implicit, i.e., done immediately before any unary operation on $q_i$. For simplicity, we assume 
that the circuit contains only unary and \cz gates, as the symmetry of \cz gates facilitates a simpler description; CNOT gates can be handled similarly (\S\ref{sec:eval}).

\subsection{\bf Selecting CEs (and EPs) to Execute Remote Gates}
The CE-selection problem is a generalization of the classical set-cover problem, as discussed below. We start with formalizing how CEs enable gate execution.

\para{Execution (Coverage) of a Remote Gate by CEs.} 
To execute a remote gate $(q_i,q_j,t)$, we can either: (i) Create a linked copy of $\qi$ ($\qj$) in the computer where $\qj$ ($\qi)$ is located, or (ii)  Create linked copies of
$\qi$ and $\qj$ in a third computer where the gate operation can be performed.
Thus, a remote gate execution can be enabled by either a single CE or a pair of
CEs. We say a single CE or a pair of CEs \emph{cover} a gate if they enable
the execution of the gate.
We also define the cost $c(M)$ of a \CE $M$ as the generation latency of the 
\EP needed for $M$. 


\para{\CE Selection (\ces) Problem.} 
Given a qubit-allocation function $\eta$, 
the \ces problem is to select a set of CEs $\mathcal{M}=\{M_1, M_2, \dots M_r\}$ such that 
\begin{itemize}
    \item Every remote gate arising from the qubit-allocation function $\eta$ is covered by CE(s) in $\mathcal{M}$.
    
    \item The total latency of the EPs corresponding to the CEs is minimized, i.e., $\sum \limits _{i=1}^r c(M_i)$ is minimized.\footnote{In general, as discussed before, the total (concurrent) generation latency of a set of EPs may {\em not} be the sum of the independent latencies. Still, we assume so for simplicity in formulating the \ces problem.}
\end{itemize}
The above \ces problem has been addressed in~\cite{g2021efficient}
for the special case when the objective is to minimize the {\em number} of CEs selected; here, we generalize their approach for the above objective of minimizing the {\em total cost} of CEs.

\para{$O(\log(n)$-Approximation {\tt Greedy-CE} Algorithm.} 
The above \ces problem can be looked upon as a weighted 
set-cover problem, wherein we need to select a minimum-weighted (minimum-cost, here) 
collection of given sets (CEs, here) to cover all the elements (remote gates, here).
However, unlike the set-cover problem, in our case, an element (remote 
gate) may be covered by a {\em combination} of two sets (CEs); this renders
the simple greedy algorithm (which picks the ``best'' set in each iteration) 
without any performance guarantee.
However, a more sophisticated {\tt Greedy-CE} algorithm that, in each iteration,
{\bf {\em picks a \underline{set} of CEs}} that covers the most number of non-yet-covered 
remote gates per unit cost of the CEs picked---can be shown to deliver 
a $O(\log(n)$-approximate solution (see Theorem~\ref{thm:ce}).
However, we need to solve the non-trivial problem 
of picking such an optimal {\em set} of CEs; fortunately, 
this can be formulated\footnote{Consider a graph over CEs as vertices, where the weight on each vertex/CE $v$ is the number of remote gates covered by $v$ by itself and the weight on edge $(u,v)$ is the number of gates covered by the pair of CEs $\{u, v\}$. However, a gate may be covered in many ways---which can lead to double counting; this can be resolved
by partitioning the graph appropriately~\cite{g2021efficient}.} and solved using a generalized {\em densest subgraph} problem, as discussed below.

\softpara{Densest Subgraph with Vertex-Costs (\dsvc) Problem.} 
Let $G=(V,E)$ be a graph where each vertex $v\in V$ has a weight $w(v)$ 
and a cost $c(v)$ associated, and each edge $e$ has a weight $w(e)$ associated. The \dsvc problem is find an induced subgraph $H$ in $G$ with 
maximum value of $(\sum_{e \in E(H)} w(e) + \sum_{v \in V(H)} w(v))/\sum_{v\in V(H)} c(v)$.
For the special case of the \dsvc problem, when each vertex has a unit cost, ~\cite{g2021efficient} proposed two algorithms: an optimal linear programming and a 2-approximate greedy approach. Both algorithms and their performance guarantees can be generalized to include arbitrary vertex costs; due to limited space, we discuss only the 2-approximate greedy approach here.
The \dsvc-{\tt Greedy} algorithm to solve the above \dsvc problem is to iteratively
remove a vertex $v$ that has the lowest value 
of $\frac{w(v)+\sum \limits_{\text{$e$ incident on 
$v$ }} w(e)}{c(v)}$. We keep track of the remaining subgraphs over the iterations and pick the best among
them (i.e., the subgraph $H$ with the maximum value of $(\sum_{e \in E(H)} w(e) + \sum_{v \in V(H)} w(v))/ \sum _{v\in V(H)} c(v)$). We can show 
that  \dsvc-{\tt Greedy} delivers a 2-approximate solution.

\begin{lem-wo-prf}
The above \dsvc-{\tt Greedy} algorithm returns a $2$-approximate solution for the \dsvc problem. 
\end{lem-wo-prf}

\begin{theorem}
The above {\tt Greedy-CE} algorithm for the \ces problem delivers a $\mathcal{O}(\log n)$-approximate solution, where $n$ is the total number of circuit gates.
\label{thm:ce}
\end{theorem}

%% file: 7b-other-gens.tex
\section{\bf Other Generalizations}
\label{sec:gen}

\para{Non-Zero Swap and Gate Latencies.}  
To incorporate non-zero swap and gate latencies, we make two changes.
(i) When generating the EPs in batches, the gates and swaps are done in parallel with the generation of EPs of the {\em  following} batch. Thus, gates and swaps do not add to the total execution time---as long as their latencies are less than the EP-batch latencies (which is expected to be largely true). (ii) We only consider batches whose total latency (including swap and gate operations) is less than the decoherence threshold $\tau$.  



\para{Fidelity Constraints.} 
The fidelity of EPs may degrade during the generation process due to the quantum
operations (decoherence can also affect fidelity, 
but we have handled it separately).
In our schemes, we use the purification technique to overcome 
fidelity degradation by requiring and creating 
multiple copies of EPs~\cite{purification} 
(instead of just one) to execute each remote gate; 
we required the number of copies to be proportional to the 
entanglement path length $l$ used to generate
the EP (as fidelity degradation is somewhat proportional to $l$).

%% file: 8-Evaluation.tex
\section{\bf Evaluation}
\label{sec:eval}

\begin{figure*}[t]
\vspace{-0.2in}
\begin{subfigure}{\textwidth}
    \centering
    \includegraphics[width=0.7\textwidth]{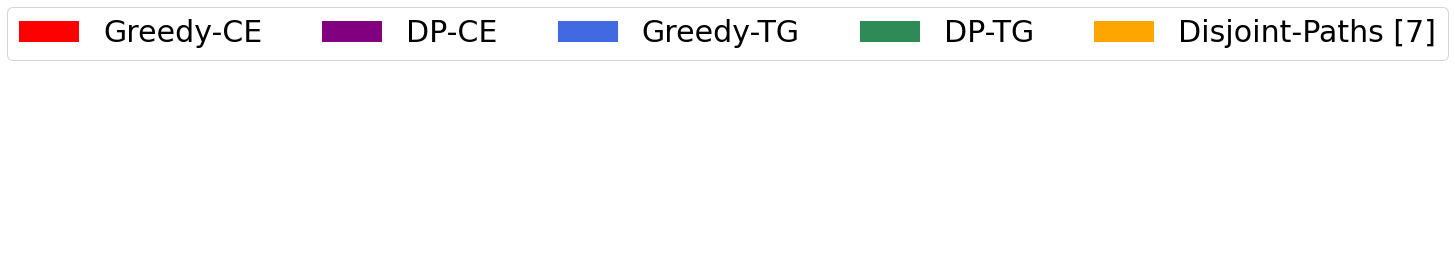}
    \vspace{-1.75cm}
\end{subfigure}
\begin{subfigure}{0.28\linewidth}
\centering
\vspace{-0.2cm}
\includegraphics[width=\textwidth]{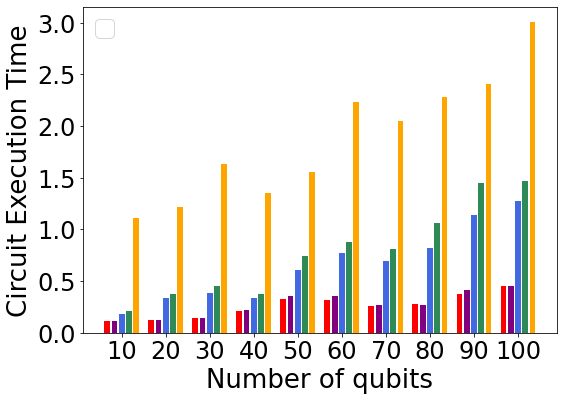}
\vspace{-0.6cm}
\captionlistentry{}
\label{fig:VaryQubits}
\end{subfigure}
\hspace{-0.3cm}
\begin{subfigure}{0.28\linewidth}
\centering
\includegraphics[width=\textwidth]{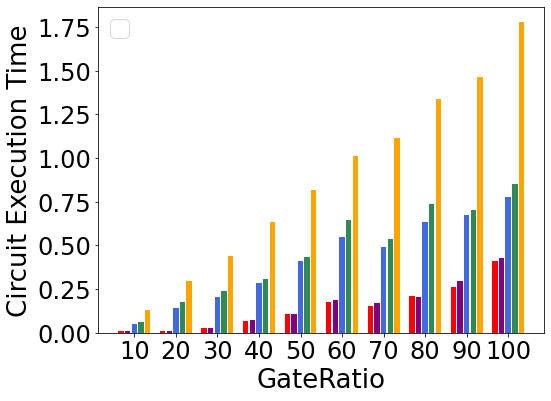}
\vspace{-0.6cm}
\captionlistentry{}
\label{fig:VarygateRatio}
\end{subfigure}
\hspace{-0.3cm}
\begin{subfigure}{0.28\linewidth}
\centering
\includegraphics[width=\textwidth]{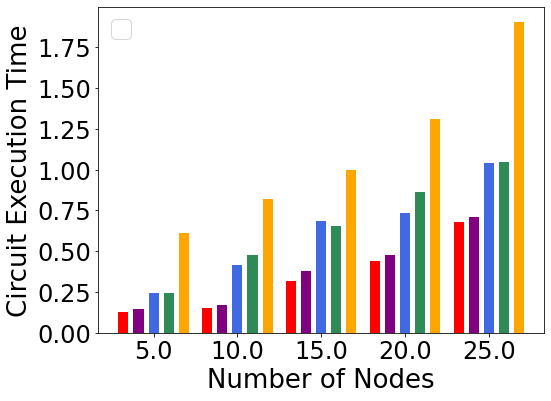}
\vspace{-0.6cm}
\captionlistentry{}
\label{fig:VaryNodes}
\end{subfigure}
\hspace{-0.3cm}
\begin{subfigure}{0.16\linewidth}
\centering
\includegraphics[width=\textwidth]{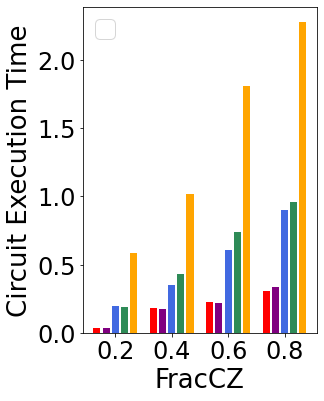}
\vspace{-0.6cm}
\captionlistentry{}
\label{fig:VaryFracCZ}
\end{subfigure}
\caption{Total execution time taken by various algorithms for varying parameters in \cz circuits.}
\label{fig:cz}
\vspace{-0.4cm}
\end{figure*}

\begin{figure*}[t]
\begin{subfigure}{0.28\linewidth}
\centering
\vspace{-0.2cm}
\includegraphics[width=\textwidth]{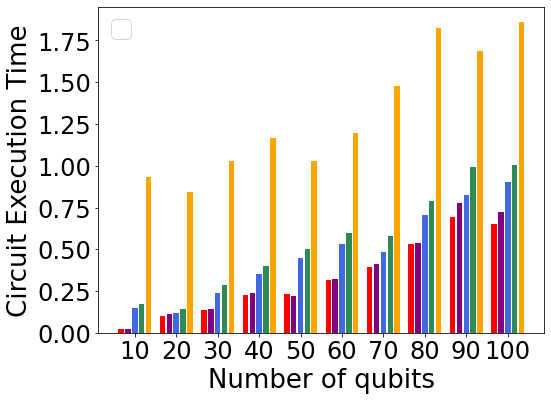}
\vspace{-0.6cm}
\captionlistentry{}
\label{fig:VaryQubitsCNOT}
\end{subfigure}
\hspace{-0.3cm}
\begin{subfigure}{0.28\linewidth}
\centering
\includegraphics[width=\textwidth]{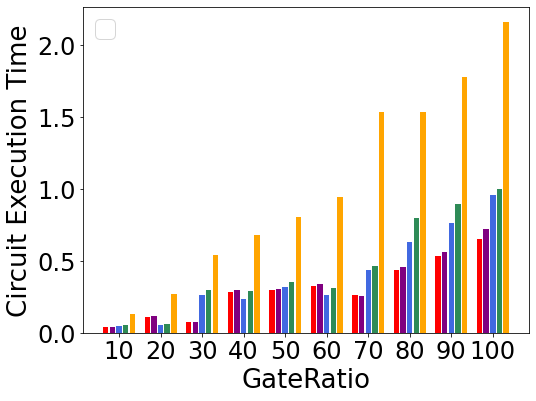}
\vspace{-0.6cm}
\captionlistentry{}
\label{fig:VarygateRatioCNOT}
\end{subfigure}
\hspace{-0.3cm}
\begin{subfigure}{0.28\linewidth}
\centering
\includegraphics[width=\textwidth]{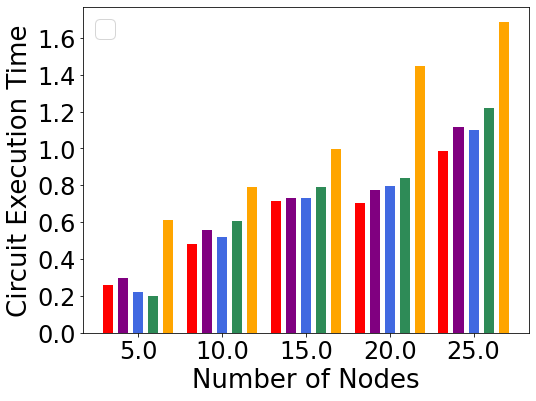}
\vspace{-0.6cm}
\captionlistentry{}
\label{fig:VaryNodesCNOT}
\end{subfigure}
\hspace{-0.3cm}
\begin{subfigure}{0.16\linewidth}
\centering
\includegraphics[width=\textwidth]{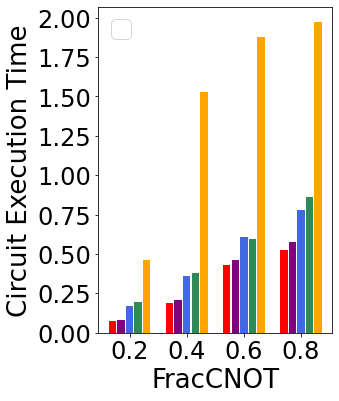}
\vspace{-0.43cm}
\captionlistentry{}
\label{fig:VaryFracCNOT}
\end{subfigure}
\caption{Total execution time taken by various algorithms for varying parameters in \CNOT circuits.}
\label{fig:cnot}
\vspace{-0.4cm}
\end{figure*}

\begin{figure*}[t]
\begin{subfigure}{0.37\linewidth}
\centering
\includegraphics[width=\textwidth]{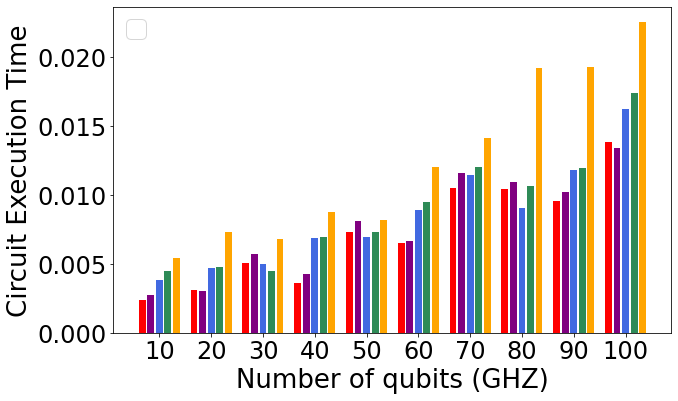}
\vspace{-0.6cm}
\captionlistentry{}
\label{fig:GHZ}
\end{subfigure}
\hspace{-0.1cm}
\begin{subfigure}{0.28\linewidth}
\centering
\includegraphics[width=\textwidth]{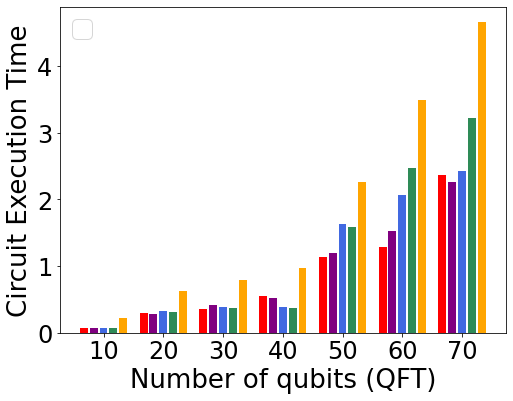}
\vspace{-0.6cm}
\captionlistentry{}
\label{fig:QFT}
\end{subfigure}
\hspace{-0.1cm}
\begin{subfigure}{0.29\linewidth}
\centering
\includegraphics[width=\textwidth]{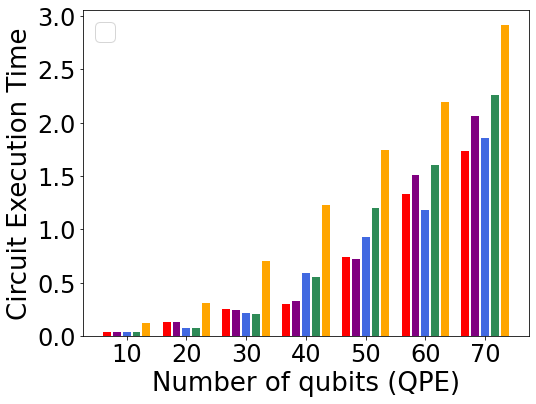}
\vspace{-0.6cm}
\captionlistentry{}
\label{fig:QPE}
\end{subfigure}
\caption{Total execution time taken by various algorithms for (\CNOT) benchmark circuits.}
\label{fig:bm}
\vspace{-0.25in}
\end{figure*}

In this section, we evaluate our algorithms over NetSquid~\cite{Netsquid}, 
a QN simulator, on random and benchmark circuits.


\para{Algorithms Compared.} We compare our techniques with the algorithm proposed in~\cite{Caleffi} (referred as \Caleffi), as it is the only work that 
considers the \dqcqr problem over general networks. 
We use our qubit-allocation scheme (\S\ref{sec:step1}) as a precursor to all the algorithms, including \Caleffi (as~\cite{Caleffi} assumes a given qubit allocation).
Our \dqcqr algorithms are named based on the execution schemes as follows: 
(i) \greedytg using telegates (\S\ref{sec:Greedy}), (ii) \DP using telegates (\S\ref{sec:DP}), (iii) \greedycat that selects near-optimal number of CEs (\S\ref{sec:ce}) and then uses the \greedyg approach over the selected CEs, and 
(iv) \DPcat, similarly.
In our evaluations, we relaxed the requirement of fixed qubit allocation by
not performing the reverse swaps; we observed this had 
minimal performance impact.




\para{Generating Random Circuits and Benchmark Circuits.}
We evaluate the techniques on random quantum circuits using the following parameters: number of qubits (default=50), number of gates per qubit (default=50), and fraction of binary gates (default=0.5).
Given the parameter values, we generate the random circuit one gate at a time. Gate operands are chosen randomly. We also evaluate on benchmark circuits corresponding to Quantum Fourier Transform (QFT), Quantum Phase Estimation (QPE), and GHZ state generation (GHZ), of various sizes obtained from the Munich Quantum Toolkit~\cite{Quetschlich_2023}.

\softpara{CNOT and CZ Circuits.}
Binary gates in the random circuits can be either all CNOT (referred to as CNOT circuits) or all CZ gates (referred to as CZ circuits). The benchmark
circuits have only CNOT and unary gates. 
The \greedycat and \DPcat schemes can be applied directly to CZ circuits; for CNOT circuits, we first convert each \CNOT gate to a \CZ gate and two unary gates before applying the CE schemes.

\para{Generating Random Networks.} We use a quantum network spread over an area of $100km \times 100km$. We use the Waxman model~\cite{waxman}, which has been used to create Internet topologies. We vary the number of network nodes from $5$ to $25$, with $10$ as the default. The total number of data memories in the network is equal to the number of circuit qubits; these data memories are randomly distributed among the network nodes.

\softpara{Network Parameters.} We use network parameter values similar to the ones used in~\cite{DPpaper}. In particular, we set the atomic-BSM probability of success and latency to be $0.4$ and $10 \mu$ seconds and the optical-BSM
probability of success to be $0.3$. We use atom-photon generation times and probability of success as $50 \mu$ sec and $0.33$, and the decoherence threshold of $1$ second.

\para{Evaluation Results.}
We evaluate the algorithms over the circuits and networks as described above.
We consider \CNOT and \CZ gate-based circuits and vary one parameter at a time while keeping the other parameters fixed to their default values mentioned above. See Figs.~\ref{fig:cz}-\ref{fig:bm}. We observe the following:
\begin{itemize}
    \item Generally, the performance of the algorithms is in the following order (best to worst): \greedycat, \DPcat, \greedytg, \DP, and \Caleffi, with our \greedycat outperforming \Caleffi~\cite{Caleffi} by up to 95\% in some cases (see below).
    \item As expected, using CEs significantly reduces execution time, especially in \cz circuits. In the CNOT circuits, the CE schemes perform worse sometimes than the telegate schemes.
\end{itemize}
We believe there are {\bf two key reasons for sometimes drastic under-performance} of \Caleffi~\cite{Caleffi}: (i) Their layering strategy is very conservative in exploiting concurrency among required EPs; (ii)  For an actual generation of EPs, they select one path for each EP's generation---focusing on maximizing the number of disjoint paths in a batch rather than generation latencies.
In contrast, our schemes use sophisticated approaches to divide the set of required EPs into batches and then use the optimal LP scheme to generate each batch together through an optimal network flow of entanglements.

\para{Runtime Overhead.} Our schemes, being polynomial-time, run in order of a few seconds for large circuits; this is a tolerable overhead, especially since these distribution strategies need to be run only one time (for a given quantum algorithm).

%% file: 9-Conclusion.tex
\section{\bf Conclusion}
In this paper, we have addressed the overarching problem of distribution of quantum circuits in quantum networks to minimize execution time, under decoherence and network resource constraints. Our future work focuses on 
developing provably near-optimal algorithms for the sub-problems addressed here, and in particular, to develop more sophisticated techniques that allow for dynamic qubit allocations.

